\def\gsim{\mathop {\vtop {\ialign {##\crcr
$\hfil \displaystyle {>}\hfil $\crcr \noalign {\kern1pt \nointerlineskip
 }
 $\,\sim$ \crcr \noalign {\kern1pt}}}}\limits}
 \def\lsim{\mathop {\vtop {\ialign {##\crcr
 $\hfil \displaystyle {<}\hfil $\crcr \noalign {\kern1pt \nointerlineskip
  }
  $\,\,\sim$ \crcr \noalign {\kern1pt}}}}\limits}
\documentclass[twocolumn]{jpsj3}

\title{Magnetic Field Effect on Crossover Temperature from Non-Fermi Liquid to Fermi Liquid Behavior in f$^2$-Impurity Systems with Crystalline-Electric-Field Singlet State Competing with Kondo-Yosida Singlet State}

\author{Shinya \textsc{Nishiyama}\thanks{E-mail address: nishiyama@blade.mp.es.osaka-u.ac.jp} and Kazumasa \textsc{Miyake} }
\inst{Division of Materials Physics, Department of Materials Engineering Science, Graduate School of Engineering Science, Osaka University, Toyonaka, Osaka 560-8531, Japan}

\recdate{July 23, 2011; accepted September 20, 2011; published online November 18, 2011}

\abst{We investigate the magnetic field dependence of the physical properties of f$^2$-configuration systems with a crystalline-electric field (CEF) singlet ground state, which gives rise to a non-Fermi liquid (NFL) fixed point due to the competition between the Kondo-Yosida singlet and CEF singlet states.
On the basis of the numerical renormalization group method, we find that the magnetic field breaks this NFL fixed point via two mechanisms: one causing the polarization of f-electrons and the other giving the ``channel'' anisotropy.
These two mechanisms induce a difference in the magnetic field dependence of the characteristic temperature $T_{\rm F}^{*}(H)$, the crossover temperature from NFL to Fermi-liquid behavior.
While the polarization of f-electrons gives $T_{\rm F}^{*}(H) \propto H^x$ ($x \simeq 2.0$), the ``channel'' anisotropy gives the $H$-independent $T_{\rm F}^{*}(H)$.
These two mechanisms cross over continuously at approximately the crossover magnetic field $H_{\rm c}$, where an anomalous $H$-dependence of $T_{\rm F}^{*}(H)$ appears.
Such $T_{\rm F}^{*}(H)$ well reproduces the NFL behaviors observed in Th$_{1-x}$U$_x$Ru$_2$Si$_2$.
We also find that the $H$-dependence of the resistivity and the magnetic susceptibility are in good agreement with the experimental results of this material.
These results suggest that the NFL behaviors observed in Th$_{1-x}$U$_x$Ru$_2$Si$_2$ can be understood if this material is located in the CEF singlet side near the critical phase boundary between the two singlet states.
}
\kword{Th$_{1-x}$U$_x$Ru$_2$Si$_2$, non-fermi liquid, magnetic field effect, f$^2$-impurity problem, crystalline-electric field effect, Kondo effect, numerical renormalization group}
\begin{document}
\maketitle
\section{Introduction}
In recent decades, non-Fermi-liquid (NFL) behaviors observed in some heavy fermion compounds and high-$T_{\rm c}$ cuprates have generated interest in issues on the quantum-critical point (QCP).
Of these NFL behaviors, those based on a single correlated impurity in systems with the f$^2$-configuration are classified into two subclasses in which the QCP is triggered by the local criticality: one is caused by the two-channel Kondo (TCK) effect due to the non-Kramers doublet state \cite{cox1987,sacramento1991,pang1991,affleck1992,sakai1996,kusunose1996,cox1998, zarand2002, yotsuhashi2002b}, and the other is caused by the competition between the crystalline-electric field (CEF) singlet and the Kondo-Yosida (K-Y) singlet states \cite{shimizu1995,yotsuhashi2002, hattori2005, nishiyama2010}.
Each of these mechanisms shows NFL behaviors below its characteristic energy scale $T_x$ because the systems flow toward an unstable fixed point.
However, in real systems, small but relevant perturbations, leading the systems away from an unstable fixed point, give rise to a finite characteristic temperature $T_{\rm F}^{*}$, the crossover temperature from NFL behavior to Fermi-liquid behavior.
Namely, these two NFL behaviors are observed in the temperature ($T$) region $T_{\rm F}^{*} \le T \le T_x$ when $T_{\rm F}^{*} \ll T_x$.\par
The NFL behaviors due to these two mechanisms are, in general, difficult to distinguish experimentally, especially in the case of U-impurity compounds, because there exists some ambiguity in determining the CEF level scheme of U ions.
Th$_{1-x}$U$_x$Ru$_2$Si$_2$ (x$\le$0.07) is one such complicated heavy fermion impurity system.
The NFL behaviors of this material are well scaled by impurity concentrations, so that many theoretical and experimental works have been carried out on the basis of these two mechanisms treating the U ion as an impurity \cite{amitsuka1994,shimizu1995,sakai1996,Marumoto1996,amitsuka2000,yotsuhashi2002}.
In fact, the NFL behaviors of Th$_{1-x}$U$_x$Ru$_2$Si$_2$, such as the $-\ln T$ divergence of both the magnetic susceptibility $\chi_{\rm imp}$ and the Sommerfeld coefficient $\gamma_{\rm imp}\equiv C_{\rm imp}/T$, $C_{\rm imp}$ being the specific heat due to the impurity, and the anomalous temperature dependence of the resistivity $\rho_{\rm imp}$, are consistent with those predicted by theories on the basis of these two mechanisms.
In the case of R$_{1-x}$U$_x$Ru$_2$Si$_2$ (R=Y and La), the degrees of NFL behaviors are less prominent.
Namely, Fermi liquid behaviors recover in the low temperature regions where Th$_{1-x}$U$_x$Ru$_2$Si$_2$ exhibits prominent NFL behaviors\cite{amitsuka2000, yokoyama2002}.
These differences can be understood from the viewpoint that the distances from the QCP are different from compound to compound.
In other words, Th$_{1-x}$U$_x$Ru$_2$Si$_2$ is assumed to be accidentally located near the QCP.\par
With the application of a magnetic field, however, there exist some aspects inconsistent with the NFL behaviors on the basis of the TCK effect even in the case of Th$_{1-x}$U$_x$Ru$_2$Si$_2$.
First, if the NFL behaviors originated from the TCK effect, the magnetic field would induce the increase in $\gamma_{\rm imp}$ due to the release of the residual entropy by lifting the degeneracy due to the doublet $\Gamma_{5}^{(2)}$ ground state of $J=4$ orbitals in tetragonal symmetry.
However, the suppression of the $-\log T$ divergence of $\gamma_{\rm imp}$ is observed in Th$_{1-x}$U$_x$Ru$_2$Si$_2$ by applying a magnetic field \cite{amitsuka2000}.
Next, it was reported that $T_{\rm F}^{*}(H)$ of Th$_{1-x}$U$_x$Ru$_2$Si$_2$ shows an anomalous magnetic field ($H$) dependence, i.e., linear in $H$ \cite{toth2010}, in contrast to the quadratic dependence expected in the TCK model\cite{sacramento1991,affleck1992,cox1998,zarand2002}.
Considering these inconsistencies, it is troublesome to argue that the NFL behaviors in Th$_{1-x}$U$_x$Ru$_2$Si$_2$ can be explained by the theory based on the TCK effect \cite{yotsuhashi2002}.
\par
In this paper, we study the magnetic field dependence of the NFL behaviors due to the competition between the K-Y singlet and the CEF singlet states in tetragonal symmetry, and discuss its applicability to the magnetic properties of Th$_{1-x}$U$_x$Ru$_2$Si$_2$.
Yotsuhashi $et$ $al$. have already discussed this problem on the basis of the two-orbital Anderson model with the ``antiferromagnetic'' Hund's rule coupling \cite{yotsuhashi2002}, the same as in the present paper.
They have shown that the logarithmic increase in $\gamma_{\rm imp}$ due to the competition between the two singlet states is suppressed by applying the magnetic field in a wide set of parameters near the unstable fixed point, which is consistent with the experimental results of Th$_{1-x}$U$_x$Ru$_2$Si$_2$ in a wide-temperature region.
Here, we also take the same CEF scheme as that in ref. \citen{yotsuhashi2002}, and investigate the $H$-dependence of the magnetic susceptibility $\chi_{\rm imp}$, the resistivity $\rho_{\rm imp}$, and the characteristic temperature $T_{\rm F}^{*}(H)$ obtained from these physical quantities.
On the basis of the Wilson numerical renormalization group (NRG) method \cite{wilson1975}, we show that the $H$-dependence of $T_{\rm F}^{*}(H)$ changes at around the crossover magnetic field $H_{\rm c}$, and that $T_{\rm F}^{*}(H)$ at $H \sim H_{\rm c}$ reproduces the anomalous behavior observed in Th$_{1-x}$U$_x$Ru$_2$Si$_2$.
Namely, the anomalous properties in Th$_{1-x}$U$_x$Ru$_2$Si$_2$ can be fully explained by the present model.
Moreover, the anomalous properties in R$_{1-x}$U$_x$Ru$_2$Si$_2$ (R=Y and La) are also consistent with our results.\par
This paper is organized as follows.
In \S2, we introduce the model Hamiltonian to discuss the competition between the two singlet states.
In \S3, the numerical result obtained by the NRG calculation of the magnetic field effect on $T_{\rm F}^{*}(H)$, the resistivity $\rho_{\rm imp}$, and the magnetic susceptibility $\chi_{\rm imp}$, are given in the cases of both singlet ground states. 
In \S4, we discuss a scaling property of the $H$-dependence of $T_{\rm F}^*(H)$ and its origin on the basis of the similarity of the unstable fixed point to the case of the TCK effect. 
In \S5, we discuss the applicability of this scenario to the experimental result of Th$_{1-x}$U$_x$Ru$_2$Si$_2$, and summarize our results in \S6.
\section{Model Hamiltonian}
\begin{figure}[t]
\begin{center}
	\includegraphics[width = 0.30\textwidth]{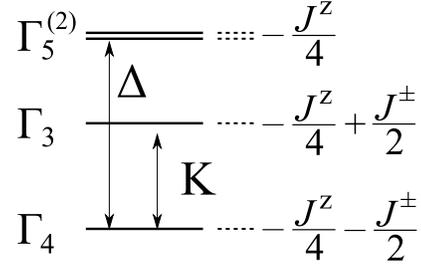}
	\caption{CEF level scheme of low-lying $f^2$ states and their eigenstates. }
	\label{fig1} 
\vspace{-8mm}
\end{center}
\end{figure}
To discuss the competition between the two singlet states, we rewrite f$^2$-states in the {\it j-j} coupling scheme using f$^1$-states in the $j=5/2$ manifold.
Here, we restrict the Hilbert space of f$^1$-states to two low-lying Kramers pairs and allot them the pseudospin states as follows\cite{yotsuhashi2002, nishiyama2010}:
\begin{align}
	\label{1a}
	\vert \Gamma_{7+}^{(2)} \rangle &= \frac{3}{\sqrt{14}} \vert  + \frac{5}{2} \rangle - \sqrt{\frac{5}{14}} \vert - \frac{3}{2} \rangle \equiv \vert \uparrow, 0\rangle, \\
	\label{1b}
	\vert \Gamma_{7-}^{(2)} \rangle &= -\frac{3}{\sqrt{14}} \vert  - \frac{5}{2} \rangle + \sqrt{\frac{5}{14}} \vert + \frac{3}{2} \rangle \equiv \vert \downarrow, 0\rangle,\\
	\label{1c}
	\vert \Gamma_{6,+} \rangle &= \vert + \frac{1}{2} \rangle \equiv \vert 0, \uparrow\rangle, \\
	\label{1d}
	\vert \Gamma_{6,-} \rangle &= \vert - \frac{1}{2} \rangle \equiv \vert 0, \downarrow\rangle.
\end{align}
The f$^2$-states are also restricted to the four low-lying CEF states in the $J=4$ manifold of tetragonal symmetry, which are written in the {\it j-j} coupling scheme within a manifold of $j=5/2$ in f$^1$-configuration as follows\cite{yotsuhashi2002, nishiyama2010}: 
\begin{align}
	\label{f2a}
	\vert \Gamma_4 \rangle &= \frac{1}{\sqrt{2}} \left( \vert  +2 \rangle - \vert -2 \rangle \right) =\frac{1}{\sqrt{2}} \left( \vert \hspace{-1.0mm} \downarrow, \uparrow \rangle - \vert \hspace{-1.0mm} \uparrow, \downarrow \rangle \right),\\
	\label{f2b}
	\vert \Gamma_3 \rangle &= \frac{1}{\sqrt{2}} \left( \vert +2 \rangle + \vert -2 \rangle \right) =\frac{1}{\sqrt{2}} \left( \vert \hspace{-1.0mm} \uparrow, \downarrow \rangle + \vert \hspace{-1.0mm} \downarrow, \uparrow \rangle \right),\\
	\label{f2c}
	\vert \Gamma_{5,+}^{(2)} \rangle &= \beta \vert +3 \rangle - \alpha \vert -1 \rangle = \vert \hspace{-1.0mm} \uparrow, \uparrow \rangle,\\ 
	\label{f2d}
	\vert \Gamma_{5,-}^{(2)} \rangle &= \beta \vert -3 \rangle - \alpha \vert +1 \rangle = \vert \hspace{-1.0mm} \downarrow, \downarrow \rangle.
\end{align}
Here, we assume the $\Gamma_4$ singlet ground state shown in Fig. \ref{fig1}, where $K$ and $\Delta$ represent the excitation energies.
With the use of the pseudospin states (\ref{1a})-(\ref{1d}), the f$^2$-level scheme is reproduced by the ``antiferromagnetic'' Hund's rule coupling\cite{yotsuhashi2002, nishiyama2010}
\begin{align}
	\label{2}
	\mathcal{H}_{\rm Hund} &= \frac{J_{\perp}}{2} \left[ S_1^{+}S_{2}^{-} +S_1^{-}S_{2}^{+} \right] + J_z S_{1}^{z}S_{2}^{z},
\end{align}
where coupling constants are defined as $J_{\perp} =K$ and $J_z = 2\Delta -K$.
$\vec{S}_m$ is a pseudospin operator of the f-electron in the Hilbert space of the f$^1$-state spanned by the orbitals $m=1$ $(\Gamma_7^{(2)})$ or $2$ $(\Gamma_6)$, and is defined as
\begin{equation}
	\vec{S}_m \equiv \frac{1}{2}f_{m\sigma}^{\dagger} \vec{\sigma}_{\sigma,\sigma^{'}} f_{m\sigma^{'}}, 
	\label{op}
\end{equation}
where $f_{m\sigma}$ is an annihilation operator of the f-electron in orbital $m$ \cite{yotsuhashi2002, nishiyama2010}.
The use of the $j-j$ coupling scheme for f$^2$-states is not necessary, in principle, for solving the present problem.
However, it makes the problem more tractable in calculations based on the Wilson NRG method. \par
Thus, the Hamiltonian is given by the two-orbital impurity Anderson model with the ``antiferromagnetic'' Hund's rule coupling as\cite{yotsuhashi2002, nishiyama2010}:
\begin{align}
	\label{3a}
	\mathcal{H} &= \mathcal{H}_{\rm c} + \mathcal{H}_{\rm hyb} + \mathcal{H}_{\rm f} + \mathcal{H}_{\rm Hund}, \\
	\label{3b}
	\mathcal{H}_{\rm c} &= \sum_{m=1,2} \sum_{\vec{k}\sigma} \varepsilon_{\vec{k}} c_{\vec{k}m\sigma}^{\dagger} c_{\vec{k}m\sigma},\\
	\label{3c}
	\mathcal{H}_{\rm hyb} &= \sum_{m=1,2} \sum_{\vec{k}\sigma} \left( V_{m} c_{\vec{k}m\sigma}^{\dagger} f_{m\sigma} + {\rm h.c.} \right),\\
	\label{3d}
	\mathcal{H}_{\rm f} &=\sum_{m\sigma}E_{fm} f_{m\sigma}^{\dagger}f_{m\sigma} + \sum_{m}U_m f_{m\uparrow}^{\dagger}f_{m\uparrow}f_{m\downarrow}^{\dagger}f_{m\downarrow},
\end{align}
where $c_{m\vec{k}\sigma}$ is the annihilation operator of a conduction electron with the wave vector $\vec{k}$ and the spin $\sigma$ hybridizing with the f-electron in the orbital $m$ with a strength $V_{m}$.
$E_{fm}$ and $U_m$ are the energy level of the f-electron and an intra-orbital Coulomb repulsion in orbital $m$, respectively, and the other Coulomb repulsion terms, like inter-orbital interaction, are implicitly included in the ``antiferromagnetic'' Hund's rule coupling (\ref{2}).\par
We consider the case when the magnetic field is applied in the $z$-direction, the $c$-axis of Th$_{1-x}$U$_x$Ru$_2$Si$_2$.
The effect of the magnetic field for f$^1$-states is taken into account through the Zeeman terms defined by 
\begin{equation}
	\mathcal{H}_{\rm Zeeman} ({\rm f}^1) = - \sum_{m} g_{m} \mu_{\rm B} S_{m}^{z} H,
	\label{mag}
\end{equation}
where the $g$-factors of orbitals 1 and 2 are $g_1 = 90/49$ and $g_2 = 6/7$, respectively.
The effects of the magnetic field for f$^n$-states ($n=2,3,4$) are calculated using $\mathcal{H}_{\rm Zeeman}({\rm f}^n)$, which is the sum of the Zeeman term $(\ref{mag})$ for each f-electron.
For example, $\langle \Gamma_{5,\pm}^{(2)} \vert \mathcal{H}_{\rm Zeeman}({\rm f}^2) \vert \Gamma_{5,\pm}^{(2)} \rangle =\pm (g_1 + g_2) \mu_{\rm B} H/2$ for the f$^2$-state $ \vert \Gamma_{5,\pm}^{(2)} \rangle$, i.e., (\ref{f2c}) and (\ref{f2d}), and $\langle \uparrow \downarrow, \uparrow \vert \mathcal{H}_{\rm Zeeman}({\rm f}^3) \vert \uparrow \downarrow, \uparrow \rangle = g_2 \mu_{\rm B} H/2$ for the f$^3$-state $\vert \uparrow \downarrow, \uparrow \rangle \equiv f_{1\uparrow}^{\dagger}f_{1 \downarrow}^{\dagger} f_{2 \uparrow}^{\dagger}\vert 0 \rangle$, where $\vert 0 \rangle$ is the vacuum state, and so on.
In the same manner, the Van Vleck contribution arising from the off-diagonal term between $\Gamma_4$ and $\Gamma_3$ in the f$^2$-singlet manifold is estimated as $\langle \Gamma_3 \vert \mathcal{H}_{\rm Zeeman} ({\rm f}^2) \vert \Gamma_4 \rangle = -(g_1 - g_2) \mu_{\rm B} H/2$.
However, this value is much smaller than that estimated in the $LS$-coupling scheme in the $J=4$ manifold, $ \langle \Gamma_4 \vert -g_{J} \mu_{\rm B} J_z H \vert \Gamma_3 \rangle = -2 g_{J} \mu_{\rm B} H_z$ with $g_{J}=4/5$, because the higher $\Gamma_7^{(1)}$ doublet state in the $j=5/2$ manifold in the f$^1$-configuration has been discarded in constructing our model Hamiltonian shown by eqs. (\ref{3a})-(\ref{3d}).
In fact, if we construct the $\Gamma_4$ and $\Gamma_3$ singlet states in the $j=5/2$ manifold as
\begin{align}
	\label{reala}
\notag	\vert \Gamma_3 \rangle &=& \frac{3}{2\sqrt{7}} \left( \vert +\frac{5}{2} \rangle \vert - \frac{1}{2}\rangle + \vert +\frac{1}{2} \rangle \vert - \frac{5}{2}\rangle \right)\\
	&&+ \frac{1}{2} \sqrt{ \frac{5}{7}} \left( \vert +\frac{3}{2} \rangle \vert + \frac{1}{2}\rangle + \vert -\frac{1}{2} \rangle \vert - \frac{3}{2}\rangle \right), \\
	\label{realb}
\notag	\vert \Gamma_4 \rangle &=& \frac{3}{2\sqrt{7}} \left( \vert +\frac{5}{2} \rangle \vert - \frac{1}{2}\rangle - \vert +\frac{1}{2} \rangle \vert - \frac{5}{2}\rangle \right) \\
	&&+ \frac{1}{2} \sqrt{ \frac{5}{7}} \left( \vert +\frac{3}{2} \rangle \vert + \frac{1}{2}\rangle - \vert -\frac{1}{2} \rangle \vert - \frac{3}{2}\rangle \right), 
\end{align}
the off-diagonal term is estimated as $\langle \Gamma_3 \vert \mathcal{H}_{\rm Zeeman} ({\rm f}^2) \vert \Gamma_4 \rangle = 2 g_j \mu_{\rm B} H$ with $g_j=6/7$, which almost coincides with the value estimated in the $LS$-coupling scheme in the $J=4$ manifold.
Thus, to take into account the Van Vleck contribution properly, we adopt the off-diagonal matrix element in the $J=4$ manifold other than the contribution of the f$^1$-based Zeeman term.\par
We transform the conduction band part of the Hamiltonian (\ref{3b}), with a logarithmic discretization parameter, $\Lambda=2.5$, into the one-dimensional semi-infinite chain model and carry out the Wilson NRG method\cite{wilson1975}.
For simplicity, we take conduction bands to be symmetric in the energy space (with an extent from $-D$ to $D$) centered at the Fermi level.
We keep the low-lying 4000 states in each iteration step.\par
\section{Characteristic Temperature $T_{\rm F}^{*}(H)$}
The Hamiltonian (\ref{3a}) has two stable fixed points.
One is the K-Y singlet fixed point (KY SFP) where the spin degree of freedom of each f-electron is screened by the conduction electrons with the same symmetry as the f-electron, leading to the phase shift in the unitarity limit as $\delta_m = \pi/2$ ($m=1,2$).
The other is the CEF singlet fixed point (CEF SFP) where two f-electrons form the singlet state due to the CEF effect, characterized by $\delta_m = 0$ ($m=1,2$).
Along the boundary of these two stable-fixed-point regions, there exists a locus of the unstable fixed points across which the ground state is interchanged.
Around this line, NFL behaviors appear at $T_{\rm F}^{*} < T < T_x= {\rm min}(T_{\rm K2},K)$, where $T_{\rm K2}$ is the lower Kondo temperature of two f-orbitals.\par
In general, $E_{fm}$ and $U_m$, the energy level and the Coulomb interaction of each f-orbital, respectively, are different.
However, for simplicity, we take the same values for each orbital, and the difference in characters of each orbital is introduced only through $V_m$.
The Kondo temperature of orbital 2 is postulated to always be lower than that of orbital 1, i.e., $T_{\rm K1} > T_{\rm K2}$, and the parameters of the Hamiltonian (\ref{3a}) are fixed as $E_{f1}= E_{f2}=-0.4, U_1=U_2=1.5, V_1=0.45$ and $V_2=0.30$ in the unit of $D$ throughout this paper.
In addition, the magnetic field $H$ is measured in the unit of $D/\mu_{\rm B}$.
In the case of $K=\Delta=0$, the Hamiltonian (\ref{3a}) reduces to two independent impurity Anderson models, where the Kondo temperatures determined by the definition of Wilson, i.e., $4T_{\rm K}\chi_{\rm imp}(T=0) = 0.413$, are $T_{\rm K1}= 4.52 \times 10^{-2}$ and $T_{\rm K2}= 3.43 \times 10^{-3}$.
In this paper, we set the CEF level splittings as $K>\Delta$, which reproduces the anisotropy of the magnetic susceptibility, $\chi_z > \chi_{\perp}$, as pointed out in ref. \citen{yotsuhashi2002}.
Moreover, we fix $\Delta=0.12$ and control the degree of the competition by varying the CEF level splitting $K$.
For the parameter set above, $K^{*} \simeq 0.0464$ gives an unstable fixed point, i.e., the ground state is the K-Y singlet for $K < K^{*}$ and the CEF singlet for $ K > K^{*}$.\par
\begin{figure*}[t]
\begin{center}
	\includegraphics[width = 0.75\textwidth]{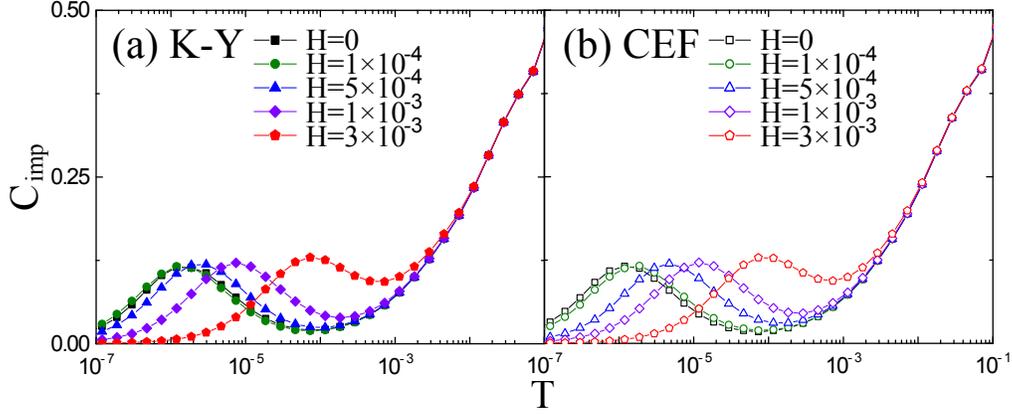}
      \caption{(Color online) Temperature dependence of the specific heat $C_{\rm imp}$ for a series of magnetic fields $H$ ($0 \le H \le 3 \times 10^{-3}$) for (a) $K=0.0440$ with $T_{\rm F}^{*} = 1.44 \times 10^{-6}$ in the KY SFP, and (b) $K=0.0488$ with $T_{\rm F}^{*} = 1.33 \times 10^{-6}$ in the CEF SFP.}
	\label{fig2} 
\end{center}
\end{figure*}

Figure \ref{fig2} shows the temperature dependence of the specific heat in two cases: $K=0.0440$ in the KY SFP region and $K=0.0488$ in the CEF SFP region.
The characteristic temperature $T_{\rm F}^{*}$ is defined as the lowest temperature at which the specific heat $C_{\rm imp}(H)=\partial S_{\rm imp}(H)/\partial \ln T$, where $S_{\rm imp}$ is the entropy due to the impurity, has a peak corresponding to the release of $\log \sqrt{2}$ entropy that characterizes the unstable fixed point \cite{yotsuhashi2002, nishiyama2010}.
For these parameters, the characteristic temperatures are obtained as $T_{\rm F}^{*} \sim 1.44 \times 10^{-6}$ in the KY SFP region and as $T_{\rm F}^{*} \sim 1.33 \times 10^{-6}$ in the CEF SFP region.
In the case of the KY SFP region, $T_{\rm F}^{*}$ slightly decreases for a magnetic field $H=1.0 \times 10^{-4}$, but increases for the other values of the magnetic field.
On the other hand, in the case of the CEF SFP region, $T_{\rm F}^{*}$ increases at all values of the magnetic field.\par
\begin{figure}[tb]
\begin{center}
	\includegraphics[width = 0.48\textwidth]{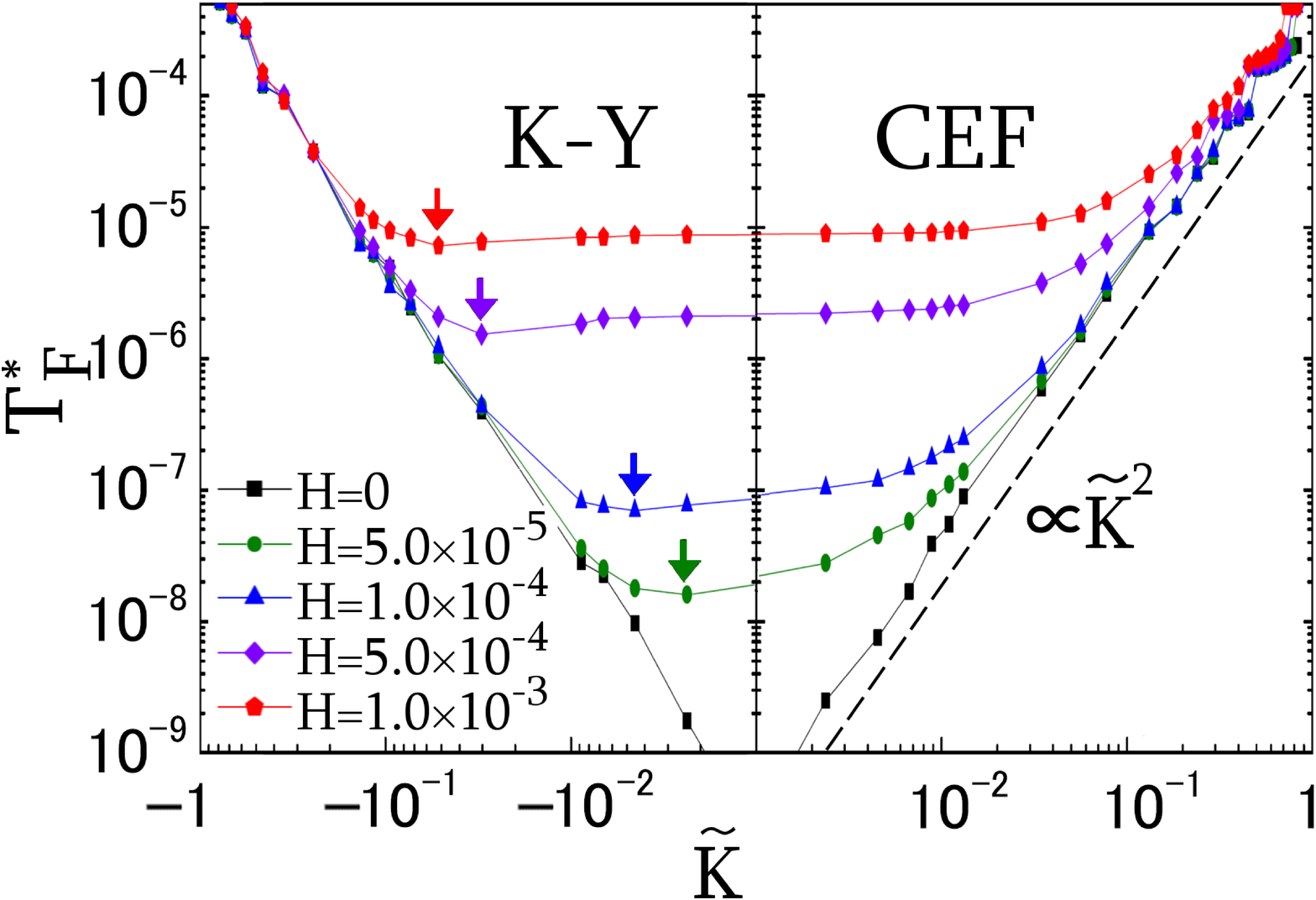}
      \caption{(Color online) Characteristic temperature $T_{\rm F}^{*}$ vs $\tilde{K} \equiv ( K-K^{*})/K^*$ for a series of magnetic fields $H$. Data points in the KY SFP are represented by closed symbols, while those in the CEF SFP are represented by open symbols. Arrows indicate the positions of the dip of $T_{\rm F}^{*}(H)$.}
	\label{fig3} 
\vspace{-8mm}
\end{center}
\end{figure}

Figure \ref{fig3} shows the $ \tilde{K} \equiv (K-K^*)/K^{*} $ dependence of $T_{\rm F}^{*}(H)$.
The characteristic temperature $T_{\rm F}^{*}$ is decreased by the competition.
One can see in Fig. \ref{fig3} that the $\tilde{K}$-dependence of $T_{\rm F}^{*}$ at $H=0$ is given by $T_{\rm F}^{*} \propto \tilde{K}^2 $ around the QCP, indicating that $T_{\rm F}^{*}$ gives a degree of deviation from the QCP.
When the magnetic field is applied, $T_{\rm F}^{*}$ increases, and the energy spectrum no longer suddenly interchanges at $K=K^{*}$ because the ground state is the mixed state between the CEF singlet and K-Y singlet states.
In the KY SFP region, there is a dip (indicated by arrow in Fig. \ref{fig3}) at which $T_{\rm F}^{*}$ takes a minimum but remains non-zero.
The energy spectrum obtained by the NRG calculation ``gradually'' crosses over between the types of the CEF and the K-Y singlet states around this dip.
Namely, it is the point where the dominant singlet state of the two singlet states interchanges.
As the magnetic field increases, this dip moves from $K=K^{*}$ to the low $K$ region, which indicates that the magnetic field increases the weight of the CEF singlet state compared with that of the K-Y singlet state.
This increase in the weight of the CEF singlet state originates from the off-diagonal term between the $\Gamma_3$ and $\Gamma_4$ f$^2$-CEF singlet states because it stabilizes the energy level of the $\Gamma_4$ CEF singlet ground state.
Hereafter, we investigate the $H$-dependence of physical quantities in two cases being close to the QCP: $K=0.0440$ with $T_{\rm F}^{*}= 1.44 \times 10^{-6}$ in the KY SFP region and $K=0.0488$ with $T_{\rm F}^{*}= 1.33 \times 10^{-6}$ in the CEF SFP region.\par

\begin{figure*}[bt]
\begin{center}
	\includegraphics[width = 0.75\textwidth]{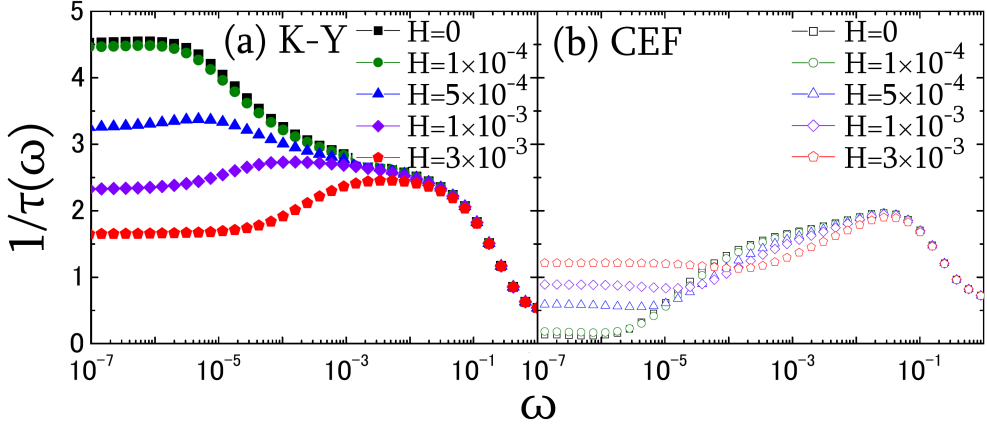}
	\caption{(Color online) Frequency dependence of the total scattering rate $1/\tau (\omega)$ for a series of magnetic fields $H$ ($0 \le H \le 3 \times 10^{-3}$) for (a) $K=0.0440$ with $T_{\rm F}^{*} = 1.44 \times 10^{-6}$ in the KY SFP, and (b) $K=0.0488$ with $T_{\rm F}^{*} = 1.33 \times 10^{-6}$ in the CEF SFP.}
	\label{fig4} 
\vspace{-8mm}
\end{center}
\end{figure*}

Figure \ref{fig4} shows the frequency dependence of the total scattering rate $1/\tau(\omega)$ at $T=0$ in two cases, i.e., $K=0.0440$ and $K=0.0488$, where $1/\tau(\omega)$ is the sum of contributions from each orbital, and the spin component $1/\tau_{m\sigma} (\omega)= 2 \pi \vert V_m \vert^2 A_{m\sigma}(\omega)$, $A_{m\sigma}(\omega)$ being the single-particle spectral function.
Data points in the KY SFP region are represented by closed symbols, while those in the CEF SFP region are represented by open symbols unless stated explicitly.
The $T$-dependence of the resistivity can be inferred from $1/\tau(\omega)$ because $\omega$ and $T$ are of the same order in the Fermi-liquid theory\cite{agd}, e.g., in the case of the single orbital Anderson model, $ 1/\tau(\omega,T) \simeq \left[ 1/\tau(0,0) \right] \left[ 1- (\omega^2 + \pi^2 T^2 )/3T_{\rm K}^2 + \cdots \right] $\cite{nozieres1974}.
Although it is not shown as figures in the present paper, $1/\tau(\omega)$ increases logarithmically in the region of $\omega \gsim T_{\rm K1}$ owing to the screening of the f-electron by conduction electrons in orbital 1 in both cases.
With decreasing $\omega$ toward $T_{\rm F}^*( \ll T_{\rm K2} < T_{\rm K1})$, $1/\tau(\omega)$ shows a logarithmic {\it increase} in the KY SFP region, but it shows a logarithmic {\it decrease} in the CEF SFP region.
Finally, the Fermi-liquid behavior is restored in both regions, i.e., $1/\tau(\omega)\propto \omega^2$, at $\omega < T_{\rm F}^*(H=0)$.\par
When the magnetic field is applied, the residual scattering rate $1/\tau_0 \equiv 1/\tau(\omega)\vert_{\omega \rightarrow 0}$ decreases in the KY SFP region, as seen in Fig. \ref{fig4}(a).
There are two origins that induce such an increase in $1/\tau_0$: one is the mixing between the K-Y and CEF singlet states in the ground state, and the other is the mixing between the $\Gamma_4$ and $\Gamma_3$ singlet states through the off-diagonal term.
As a result, the magnetic field leads to the polarization of each f-electron. These magnetic moments make a singlet state, leading to the reduction in the phase shift.
On the other hand, in the CEF SFP region, $1/\tau_0$ increases because the CEF-type ground state is polarized, and its magnetic moment scatters off conduction electrons, leading to an increase in the phase shift and $1/\tau(\omega)$ at $\omega < T_{\rm F}^{*}(H=0)$.
Because the weight of the CEF singlet state in the ground state markedly increases compared with that of the K-Y singlet state around the QCP, $1/\tau(\omega)$ in the KY SFP region shows the same $T$-dependence as in the CEF SFP region under a high magnetic field. 
After all, the $T$-dependence of the resistivity $\rho_{\rm imp}$ due to the impurity scattering is essentially given by that of $1/\tau(\omega=T)$.\par
\begin{figure*}[t]
\begin{center}
	\includegraphics[width = 0.75\textwidth]{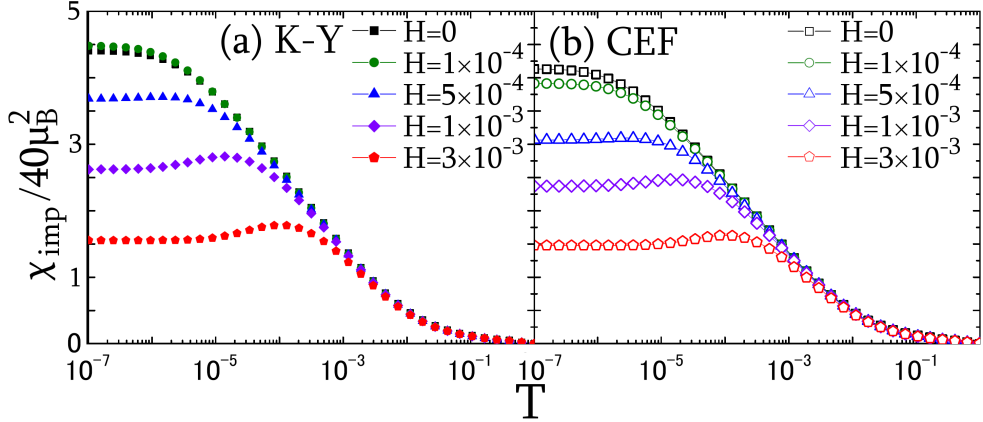}
	\caption{(Color online) Temperature dependence of the susceptibility $\chi_{\rm imp}$ for a series of magnetic fields $H$ ($0 \le H \le 3 \times 10^{-3}$) for (a) $K=0.0440$ with $T_{\rm F}^{*}=1.44 \times 10^{-6}$ in the KY SFP and for (b) $K=0.0488$ with $T_{\rm F}^{*}=1.33 \times 10^{-6}$ in the CEF SFP.}
	\label{fig5} 
\vspace{-8mm}
\end{center}
\end{figure*}

The magnetic susceptibilities $\chi_{\rm imp} \equiv \partial M/ \partial H$ are shown in Fig. \ref{fig5} in these two cases.
The magnetization $M$ consists of $M_1$ [arising from the Zeeman term $\mathcal{H}_{\rm Zeeman}({\rm f}^n)$] and $M_2$ [arising from the Van Vleck term in the f$^2$ configuration, eqs. (\ref{f2a}) and (\ref{f2b})]. 
$M_1$ is given as the thermal average of the magnetic moment $m$, which is calculated as  $m= \pm (g_1 + g_2)\mu_{\rm B}/2$ for f$^2$-state $ \vert \Gamma_{5,\pm}^{(2)} \rangle$, i.e., eqs. (\ref{f2c}) and (\ref{f2d}), and $m= \mu_{\rm B} g_2/2$ for f$^3$-state $\vert \uparrow \downarrow, \uparrow \rangle$, and so on.
On the other hand, $M_2$ is given by the effect of the off-diagonal element of the magnetization between the f$^2$-CEF singlet states $\Gamma_3$, i.e., eq. (\ref{f2b}), and $\Gamma_4$, i.e., eq. (\ref{f2a}).
In both cases, $\chi_{\rm imp}(T)$ shows the logarithmic $T$-dependence at approximately $T \sim T_{\rm K1}$ and $ T_{\rm F}^{*}< T < {\rm min}(T_{\rm K2},K)$.
The magnetic field reduces the coefficient of the $-\log T$ term at $T_{\rm F}^{*}< T < {\rm min}(T_{\rm K2},K)$ and the Van Vleck contribution.
In particular, these reductions in the KY SFP region are smaller than those in the CEF SFP region.
The origin of this phenomenon is the interchange of the weight of the two singlet states in the ground state.
At higher magnetic fields, in both cases, a broad peak appears at $T \sim T_{\rm F}^{*}(H)$, where $T_{\rm F}^{*}$ is obtained from $C_{\rm imp}$, as mentioned above. \par
\section{Scaling Behavior of Characteristic Temperature $T_{\rm F}^{*}(H)$}
\begin{figure}[ht]
\begin{center}
	\includegraphics[width = 0.43\textwidth]{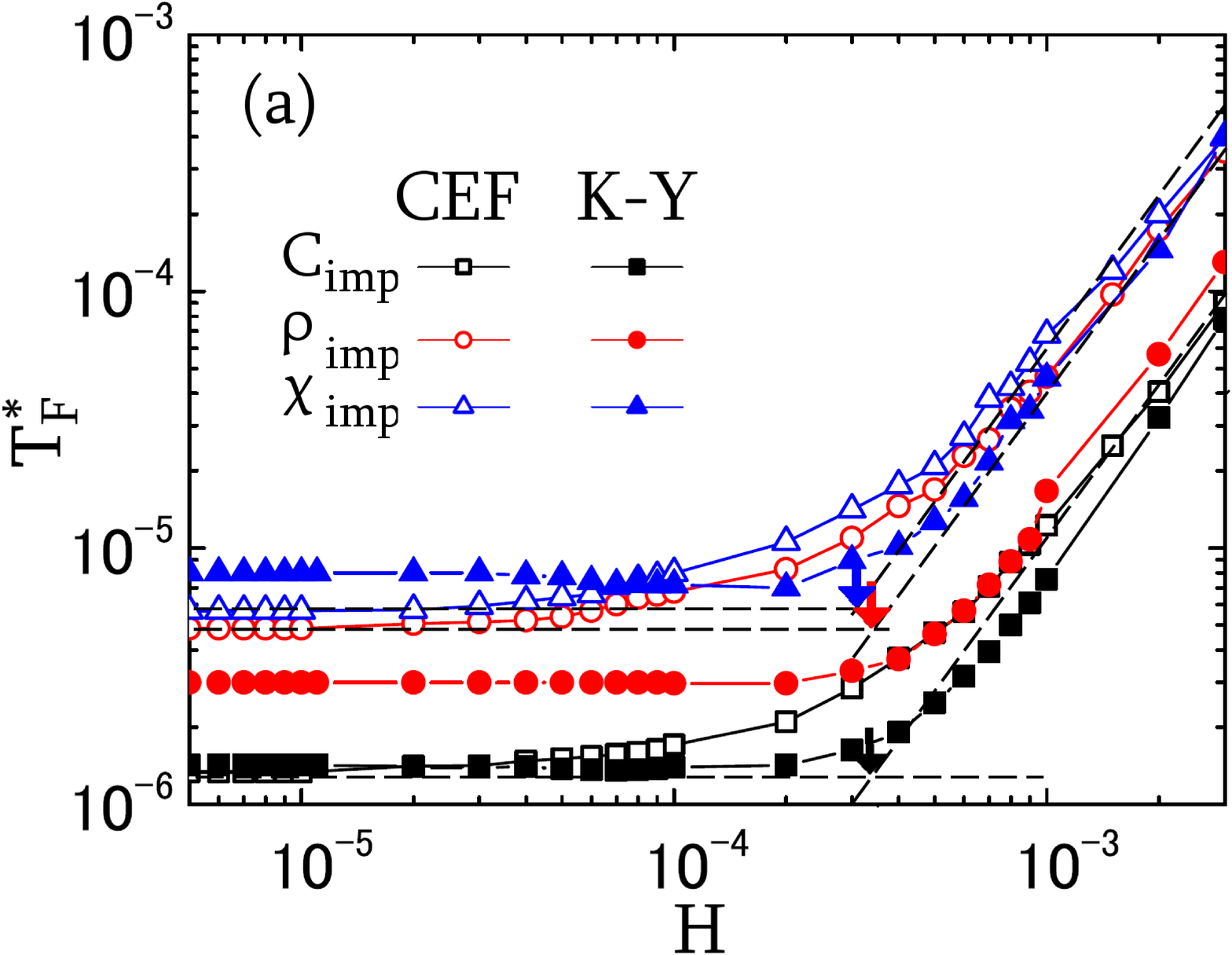}
	\includegraphics[width = 0.43\textwidth]{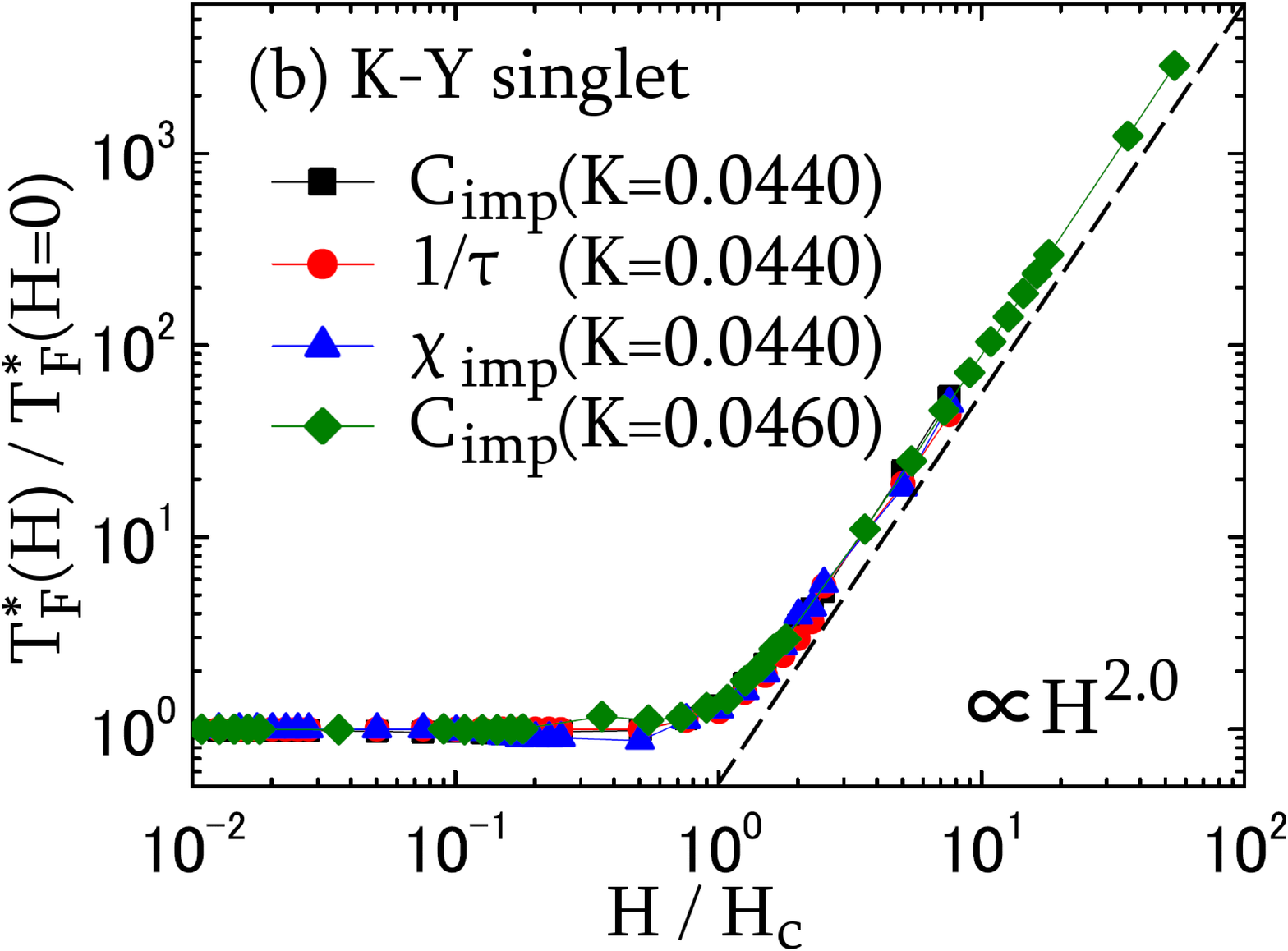}
	\includegraphics[width = 0.43\textwidth]{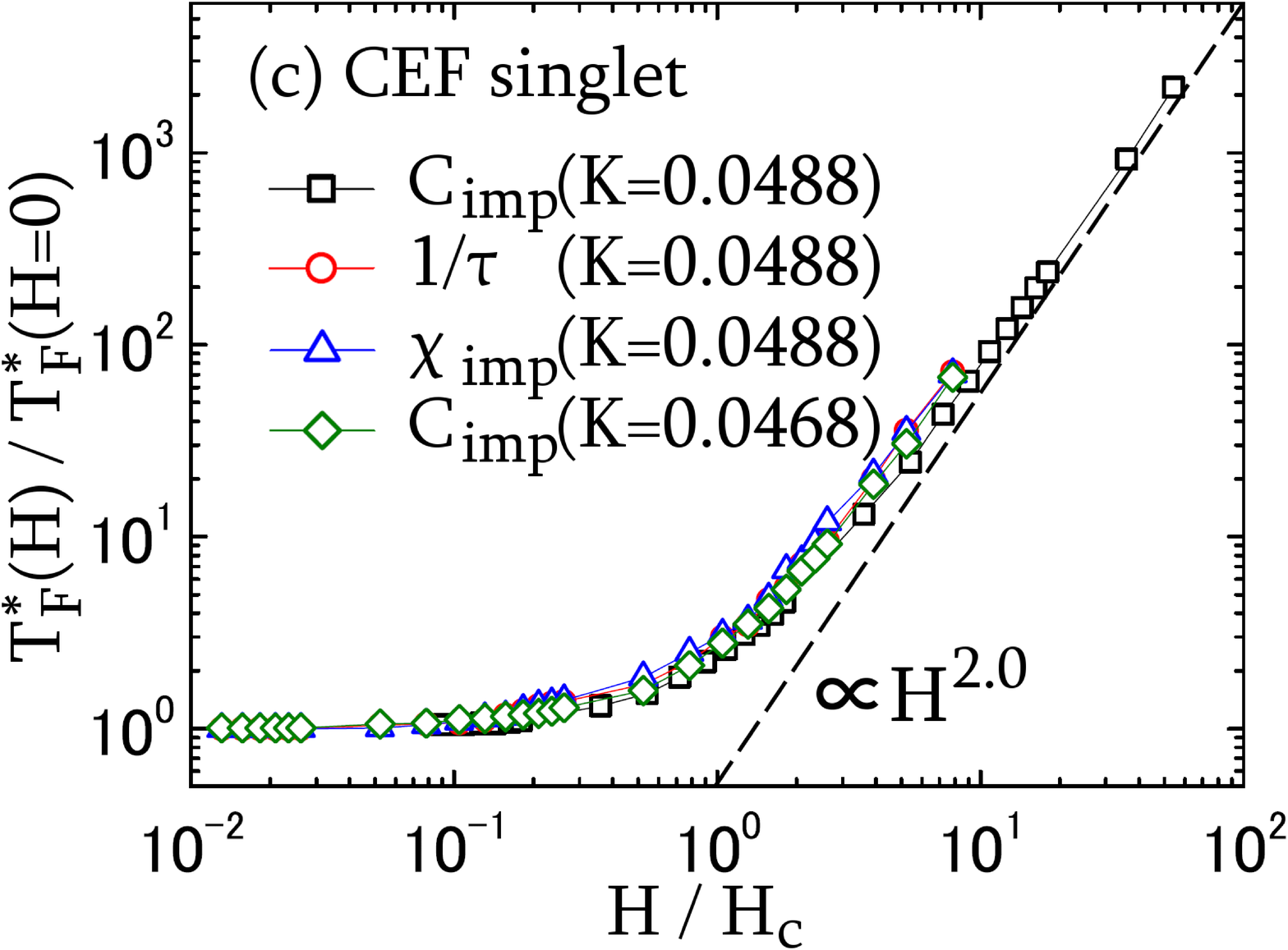}
	\caption{(Color online) (a) Characteristic temperature $T_{\rm F}^{*}(H)$ vs magnetic field $H$ in the two cases with CEF (for $K=0.0488$ and $\tilde{K}=0.05172$ shown by open symbols) and K-Y (for $K=0.0440$ and $\tilde{K}=-0.05172$ shown by closed symbols) singlet fixed points.
	The crossover magnetic fields $H_{\rm c}$'s are shown by down arrows in the case of the CEF singlet fixed point.
	(b) and (c) are scaling plots of $T_{\rm F}^{*}(H)/T_{\rm F}^{*}(H=0)$ vs $H/H_{\rm c}$ for the KY SFP and CEF SFP, respectively.}
	\label{fig6} 
\vspace{-8mm}
\end{center}
\end{figure}

From the data of the $T$ or $\omega$ dependence of $C_{\rm imp}$, $\chi_{\rm imp}$, and $1/\tau$ under the magnetic field $H$, we obtain the $H$-dependence of $T_{\rm F}^{*}(H)$, as shown in Fig. \ref{fig6}(a), where the $T_{\rm F}^{*}(H)$ of $1/\tau$ and $\chi_{\rm imp}$ are defined as the temperature at which the logarithmic $T$-dependence stops with decreasing $\omega$ and $T$ toward 0.
We define the crossover magnetic field $H_{\rm c}$ as the intersection point of linear fits (dotted lines) on the log-log plot for high and low magnetic field regions in the CEF SFP region ($K=0.0488$ and $\tilde{K}=0.05172$), as shown in Fig. \ref{fig6}(a). 
It is remarkable that the thus-determined $H_{\rm c}$'s for three different physical quantities, i.e., $C_{\rm imp}$, $\chi_{\rm imp}$, and $1/\tau$, almost coincide with each other, as shown by arrows in Fig.\ref{fig6}(a), giving a solid basis for defining the crossover magnetic field $H_{\rm c}$.
In the KY SFP region ($K=0.0440$ and $\tilde{K}=-0.05172$), the $H_{\rm c}$'s are similarly defined, although $T_{\rm F}^{*}$ obtained from $\chi_{\rm imp}$ exhibits a tiny dip near $H=H_{\rm c}$.
Linear fits are not shown for presentation clarity.
The crossover magnetic field $H_{\rm c}$ so determined for $C_{\rm imp}$, $1/\tau$, and $\chi_{\rm imp}$ almost coincide again.
These $H_{\rm c}$'s in the two singlet-fixed-point regions almost coincide with each other because the absolute values of $\tilde{K}$ for these two parameters are almost the same, while $H_{\rm c}$ depends crucially on $\tilde{K} \equiv (K-K^{*})/K^{*}$, the deviation from the QCP.
The normalized characteristic temperature, $T_{\rm F}^{*}(H)/T_{\rm F}^{*}(H=0)$, is shown in Figs. \ref{fig6}(b) and \ref{fig6}(c) as a function of the normalized magnetic field $H/H_{\rm c}$ from the three physical quantities mentioned above.
Figure \ref{fig6}(b) shows the results for the KY SFP with parameters $K=0.0440$ giving $T_{\rm F}^{*}(H=0)=1.44 \times 10^{-6}$ and $K=0.0460$ giving $T_{\rm F}^{*}(H=0) = 2.80 \times 10^{-8}$, and Fig. \ref{fig6}(c) shows those for the CEF SFP with parameters $K=0.0488$ giving $T_{\rm F}^{*}(H=0)=1.33 \times 10^{-6}$ and $K=0.0468$ giving $T_{\rm F}^{*}(H=0)= 3.79 \times 10^{-8}$.
These four lines exhibit a good scaling property in both cases, which indicates the following two important facts.\par
First, the $T_{\rm F}^{*}(H)$'s of the three quantities $C_{\rm imp}$, $1/\tau$, and $\chi_{\rm imp}$ exhibit qualitatively the same behaviors, while they are qualitatively different.
Second, the scaling property holds in both the CEF SFP and KY SFP regions even if the degree of the deviation from the QCP is different.
Note that the shapes of the normalized plots for the two stable fixed points are different from each other. 
In the region $H \ll H_{\rm c}(\tilde{K})$, $T_{\rm F}^{*}(H)$ is independent of $H$ and $T_{\rm F}^{*}(H)$ is robust particularly against the low magnetic field, so that the three physical quantities discussed above are not affected appreciably.
We previously found this robustness of $T_{\rm F}^{*}(H)$ against $H$, as reported in ref. \citen{nishiyama2010}.
On the other hand, all the $T_{\rm F}^{*}(H)$'s show the $H$-dependent form as $T_{\rm F}^{*} (H) \propto H^{x}$ in the region $H_{\rm c} (\tilde{K}) \ll H < {\rm min}(T_{\rm K2},K)$.
The exponent $x$ is estimated to be $x \simeq 2.0$ in both the CEF and KY SFP regions.
These two regions continuously cross over at approximately $H \simeq H_{\rm c}(\tilde{K})$.
Of course, it is possible that $T_{\rm F}^{*}(H)$ can be fitted as $T_{\rm F}^{*}(H) \propto H$ in a very narrow region of the magnetic field near $H = H_{\rm c}$, especially in the CEF SFP region, as shown in Fig. \ref{fig6}(c).
However, such a scaling behavior should be regarded as that of a crossover, but not an asymptotic anomalous behavior.\par
Such a $H$-dependence of $T_{\rm F}^* (H)$ can be understood by considering the similarity of the unstable fixed point to that of the TCK effect.
In the case of the TCK model, there are two origins that break the unstable fixed point, the magnetic field that polarizes the local spin leading to the ``unusual'' Fermi-liquid fixed point characterized by the energy scale $T_{\rm F}^{*} \propto H^2/T_{\rm K}$ \cite{zarand2002}, and the channel anisotropy of the exchange interaction that leads to the Fermi-liquid fixed point \cite{pang1991, affleck1992, yotsuhashi2002b}.
Indeed, the Hamiltonian (\ref{3a}) can be regarded as the TCK model below $T_{\rm K1}$ because the f-electron in orbital 2 interacts with two ``conduction'' electron channels: one is the conduction electrons in orbital 2, and the other is a complex of conduction electrons and the f-electron in orbital 1 that is screened by the conduction electrons with the same symmetry as the f-electron in orbital 1\cite{yotsuhashi2002, nishiyama2010}. 
In the present model, these two types of ``conduction'' electron serve as channels.
The change in the energy difference between the two singlet states, which is induced by the magnetic field, as mentioned above, affects the coupling constants between the f-electron in orbital 2 and the ``conduction'' electrons.
Namely, these two coupling constants exhibit magnetic field dependences.
Thus, the ``channel'' anisotropy for the two types of ``conduction'' electron is induced by the magnetic field, and the system goes to the Fermi-liquid fixed point, even though the ground state is a mixture of the two singlet states.
In other words, in the present model, the magnetic field breaks the unstable fixed point via two mechanisms, the polarization of f-electrons and ``channel'' anisotropy.\par 
In the region $H \ll H_{\rm c} $, the system flows into the Fermi-liquid fixed point induced by the ``channel'' anisotropy.
In the CEF SFP region, $T_{\rm F}^{*}$ shows little change against a low $H$.
However, in the KY SFP region, $T_{\rm F}^{*}$ slightly decreases as $H$ increases, which corresponds to the dip in Fig. \ref{fig3}.
Namely, the weight of the CEF singlet state in the ground state increases compared with that of the K-Y singlet state, and the dominant singlet state of the two singlet states interchanges at $H=H_{\rm c}$.
On the other hand, in the region $ H_{\rm c} \ll H$, the magnetic field induces the Fermi-liquid fixed point by the polarization of f-electrons, as in the case of the TCK effect, because $T_{\rm F}^{*}(H)$ is characterized by $H^2$, as in the case of the TCK effect under the magnetic field.
These two effects compete with each other at approximately $H \sim H_{\rm c}(\tilde{K})$, giving the crossover between the two regions.\par
The exponent of $H$ in $T_{\rm F}^{*}(H)$ asymptotically approaches $2.0$ in the high magnetic field region $H \gg H_{\rm c}$ in both SFP regions, as shown in Figs. \ref{fig6}(b) and \ref{fig6}(c).
However, the magnetic field necessary to reach the $T_{\rm F}^{*} \propto H^2$ behavior in the CEF SFP region is higher than that in the KY SFP region.
This difference stems from the existence of the $\Gamma_3$ excited CEF singlet state, which gives an additional magnetic field dependence for the $\Gamma_4$ CEF singlet ground state through the off-diagonal term between these two CEF singlet states.
This is verified by a NRG calculation that the exponent of $H$ in the CEF SFP region readily comes close to $2.0$ at $H > H_{\rm c}$, as in the KY SFP region, if we discard the off-diagonal term between f$^2$-CEF singlet states by the magnetic field, although no explicit result is shown here.\par
\section{Comparison with Experiment on Th$_{1-x}$U$_x$Ru$_2$Si$_2$}
\begin{figure}[t]
\begin{center}
	\includegraphics[width = 0.45\textwidth]{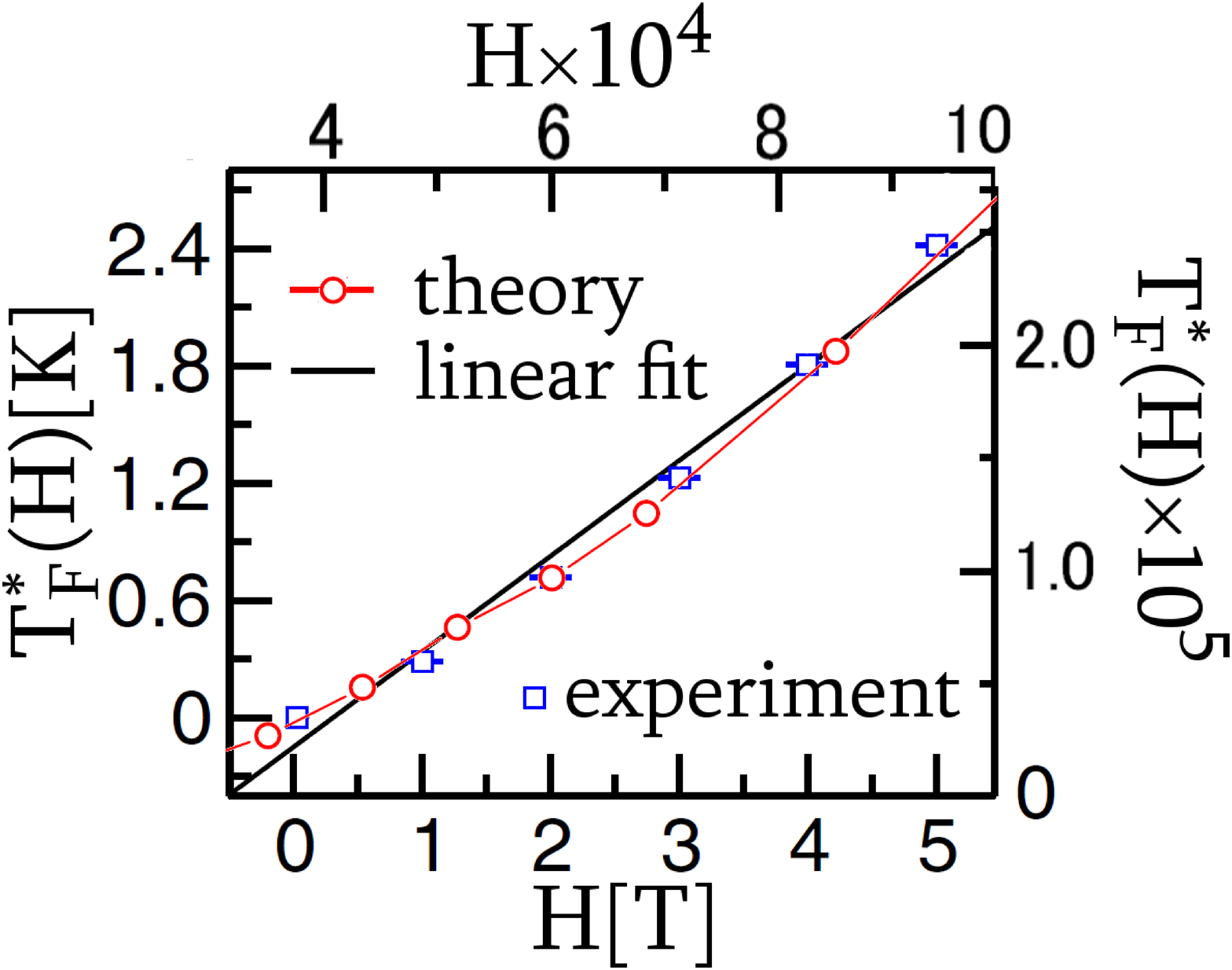}
	\caption{(Color online) Comparison between theoretical and experimental results of $T_{\rm F}^{*}(H)$.
	The red $\circ$ symbols are for the $H$-dependence of $T_{\rm F}^{*}(H)$ (upper and right scales) obtained theoretically from $C_{\rm imp}$ for $K=0.0488$.
	The blue $\Box$ symbols are for the $H$-dependence $T_{\rm F}^{*}(H)$ (lower and left scales) observed in Th$_{1-x}$U$_x$Ru$_2$Si$_2$ for the resistivity $\rho_{\rm imp}$, which was scaled linearly by T$\acute{\rm o}$th $et$ $al$. in ref. \citen{toth2010}. }
	\label{fig7} 
	\includegraphics[width = 0.40\textwidth]{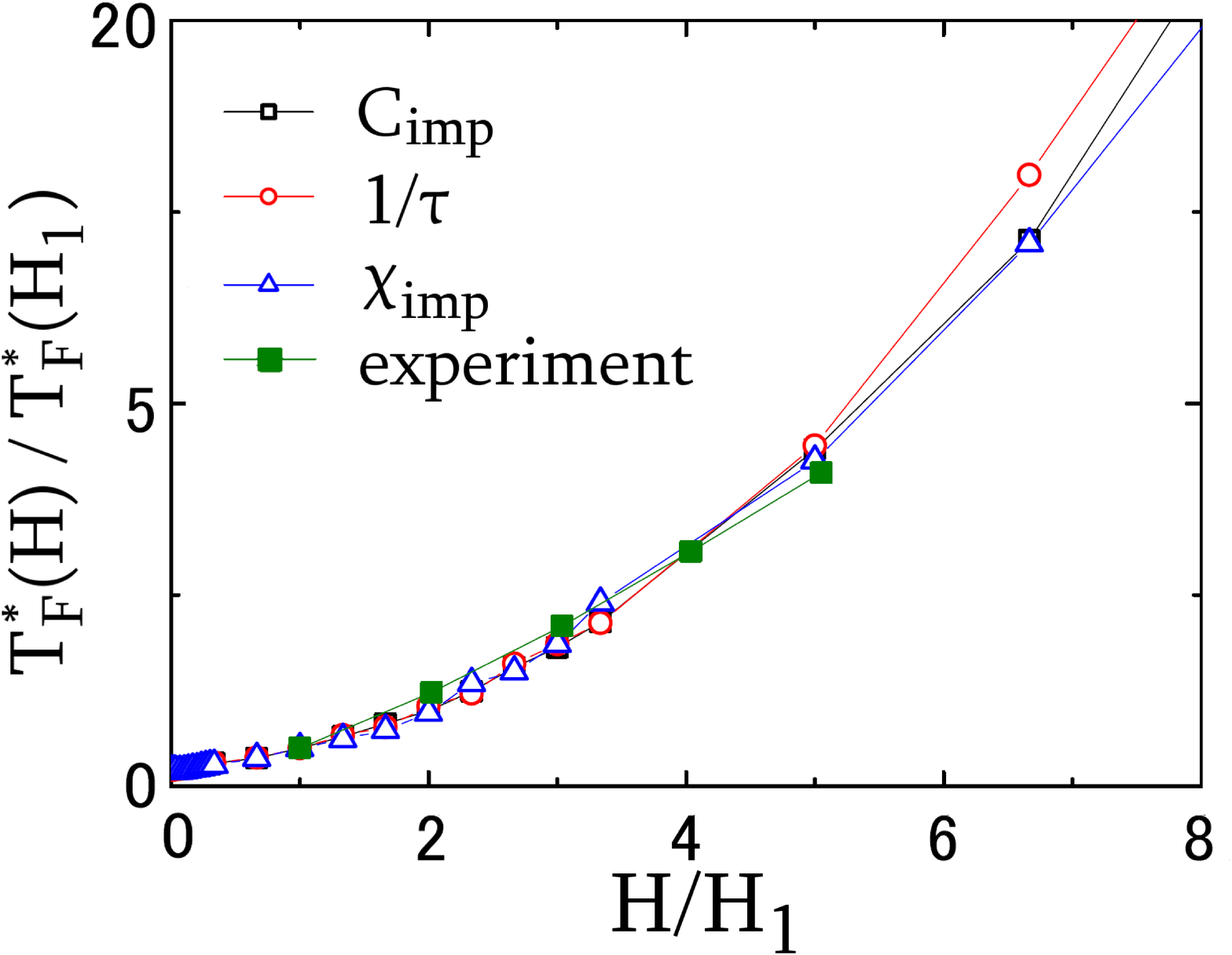}
	\caption{(Color online) $H/H_1$ vs $T_{\rm F}^{*}(H)/T_{\rm F}^{*}(H_1)$, normalized characteristic temperature, for theoretical and experimental results.
      Those obtained from $C_{\rm imp}$ and $\rho_{\rm imp}$ are normalized by the value at $H_1 = 3.0 \times 10^{-4}$, while the experimental result is normalized by the value at $H_1 \simeq 1$[T].}
	\label{fig8} 
\vspace{-8mm}
\end{center}
\end{figure}

In the CEF SFP region near the QCP, $1/\tau(\omega)$ shows a behavior consistent with the results of $\rho_{\rm imp}(T)$ in Th$_{1-x}$U$_x$Ru$_2$Si$_2$. 
Namely, $1/\tau(\omega)$ exhibits a $\log \omega$-like decrease toward $T_{\rm F}^{*}$ and increases as $H$ increases corresponding to the positive magnetic resistance.
Note that $T_{\rm K1}$ is considered to be much higher than $T_{\rm K2}$ because the $\log T$-like increase in the resistivity at approximately $T \simeq T_{\rm K1}$ is not observed in this material\cite{amitsuka1994, amitsuka2000}.
This result is consistent with that obtained by Yotsuhashi $et$ $al$. who showed that $1/\tau(\omega)$, $\chi_{\rm imp}(T)$, and the $H$-dependence of $\gamma_{\rm imp}$ reproduces these physical quantities observed in Th$_{1-x}$U$_x$Ru$_2$Si$_2$.
A more remarkable finding is that the $H$-dependence of $T_{\rm F}^{*}(H)$ at approximately $H \sim H_{\rm c}(\tilde{K})$ for $K=0.0488$ in the CEF SFP region reproduces the $H$-dependence of $T_{\rm F}^{*}(H)$ observed in Th$_{1-x}$U$_x$Ru$_2$Si$_2$ analyzed from the resistivity \cite{toth2010}.
As shown in Figs. \ref{fig7} and \ref{fig8}, our theoretical results reproduce almost perfectly the experimental observation on the normalized magnetic field $H/H_1$ dependence of $T_{\rm F}^{*}(H)$ normalized by $T_{\rm F}^{*}(H_1)$, where $H_1=3.0 \times 10^{-4} D/\mu_{\rm B}$ ($D$ being half the bandwidth of conduction electrons) for our theoretical result and $H_1 = 1$[T] for experimental result in ref. \citen{toth2010}.
Namely, our scaling plot is in good agreement with that of the experimental result of Th$_{1-x}$U$_x$Ru$_2$Si$_2$ \cite{toth2010}. \par
It is also emphasized that our theoretical analysis strongly suggests that Th$_{1-x}$U$_x$Ru$_2$Si$_2$ is located in the CEF singlet side near the critical phase boundary between the KY SFP and CEF SFP regions.
On the other hand, it is emphasized in ref. \citen{toth2010} that $T_{\rm F}^{*}(H)$ is proportional to the magnetic field $H$, especially in the analysis of the $T$-dependence of $\chi_{\rm imp}$. 
However, it is apparent that the linear fit used in ref. \citen{toth2010} fails to reproduce the experimental results for the $T_{\rm F}^{*}(H)$ obtained from the resistivity $\rho_{\rm imp}$.
The statement in ref. \citen{toth2010} stemmed from the analysis of $\chi_{\rm imp}$ assuming that the coefficient of the $-\log T$ term in $\chi_{\rm imp}$ were independent of $H$.
Experimentally, however, magnetic fields up to 5 Tesla seem to change this coefficient and markedly reduce the Van Vleck contribution to $\chi_{\rm imp}$\cite{toth2010}.\par
In the case of R$_{1-x}$U$_{x}$Ru$_2$Si$_2$ (R= La and Y), the experimental results show that the $T_{\rm F}^{*}$'s of these material are higher than that in the case of R=Th\cite{amitsuka2000, yokoyama2002}.
This indicates that parameter sets of these materials may be located more distant from the QCP than that of Th$_{1-x}$U$_x$Ru$_2$Si$_2$.
Our theoretical result predicts that pressure may induce the transition from the CEF SFP region to the KY SFP region, giving rise to a marked increase in $1/\tau_0$.\par
In the present paper, we take the same CEF level scheme as that discussed in ref. \citen{yotsuhashi2002}, because such a level scheme can reproduce the experimental results in R$_{1-x}$U$_x$Ru$_2$Si$_2$ (R=Th, Y and La). 
However, even if the low-lying CEF scheme is $\Gamma_4$-$\Gamma_5^{(2)}$-$\Gamma_3$, we have verified that similar NFL behaviors and the magnetic field dependence of $T_{\rm F}^*(H)$ occur although the results are not shown in the present paper. 
This indicates that similar NFL behaviors would be obtained if there exist a CEF singlet ground state and a strong hybridization between conduction electrons and the f-electron, namely, details of the CEF scheme would not be essential matters.
Note that CEF states with the $\Gamma_2$ singlet ground state, proposed as a plausible candidate for the ``Hidden Order'' state of URu$_2$Si$_2$ \cite{Haule2009, Kusunose2011}, would exhibit the local non-Fermi liquid behaviors discussed in the present paper.
The actual calculation in those CEF level schemes is left for future study.\par
Quite recently, it is argued that the $H$-dependence of NFL behaviors observed in Th$_{1-x}$U$_x$Ru$_2$Si$_2$ can be understood within the TCK model if some conditions would be satisfied in the CEF level scheme adopted in refs. \citen{Haule2009} and \citen{Kusunose2011}.
A crucial difference between the result in ref. \citen{toth2011} and our present one is the magnetic-field dependence of $T_{\rm F}^{*}$ in the high magnetic field region $H>5$ Tesla.
Namely, there exists a region of a magnetic field where $T_{\rm F}^{*} \propto H^2$ in the high magnetic field region in our result that is consistent with experiments as shown in Fig. \ref{fig7}, while there exists no such a region of a magnetic field in the scenario of ref. \citen{toth2011}.
\section{Conclusions}
We have investigated the magnetic field ($H$) dependence of the NFL behaviors arising from the competition between the CEF and K-Y singlet states in tetragonal symmetry.
We have found that the characteristic temperature $T_{\rm F}^{*}(H)$, below which the Fermi liquid behavior recovers, changes its $H$-dependence at approximately the critical magnetic field $H_{\rm c}$.
While $T_{\rm F}^{*}$ is not affected by $H$ in the region $H \ll H_{\rm c}$, it is expressed as $T_{\rm F}^{*} \propto H^x$ ($x \simeq 2.0)$ in the region $H \gg H_{\rm c}$.
For both high and low magnetic field regions, such a behavior of $T_{\rm F}^*(H)$ follows the scaling form even if the degree of the deviation from the QCP is different.
We have also found that $T_{\rm F}^{*}(H)$ shows an anomalous $H$-dependence at approximately $H \sim H_{\rm c}$, which is in good agreement with that observed in Th$_{1-x}$U$_x$Ru$_2$Si$_2$.
\section*{Acknowledgements}
We are grateful to H. Amitsuka and P. Chandra for stimulating discussions on their scaling analysis of Th$_{1-x}$U$_x$Ru$_2$Si$_2$ on the occasion of SCES 2010 at Santa Fe, which led our attention to the present problem.
This work is supported in part by a Grant-in-Aid for Scientific Research on Innovative Area ``Heavy Electrons'' (No.20102008) and a Grant-in-Aid for Specially Promoted Research (No. 20001004) from the Ministry of Education, Culture, Sports, Science and Technology. 
One of us (S.N.) is supported by the Global COE program (G10) from The Japan Society for the Promotion of Science.


\begin{thebibliography}{99}
\bibitem{nozieres1974}
		P. Nozi$\grave{\rm e}$res: J. Low Temp. Phys. {\bf 17} (1974) 31; Note that the coefficient of the $T^2$ term in the resistivity $\rho_{\rm imp}$ is larger than that of $1/\tau$ by three times, as discussed in this reference.
\end{thebibliography}
\end{document}